\documentclass[aps,preprint,superscriptaddress, amsmath,amssymb,floatfix,longbibliography]{revtex4-1} 
\usepackage{bm, comment, graphicx, here, amsmath, color, mathrsfs}
\usepackage[normalem]{ulem}
\usepackage[colorlinks=True, allcolors=blue]{hyperref}
\usepackage{balance, setspace, float, titlesec, array}
\usepackage{ulem}
\usepackage{siunitx}
\titlespacing*{\section}{0pt}{1ex}{0ex}
\titlespacing*{\subsection}{0pt}{1ex}{0ex}

\usepackage{xcolor}

\usepackage{soul}
\soulregister\cite7
\soulregister\ref7
\soulregister\onlinecite7

\begin{document}
\title[]{How Machine Learning Predicts Fluid Densities under Nanoconfinement}

\author{Yuanhao Li}
\email{yuanhaol@alumni.cmu.edu}  
\affiliation{Department of Chemistry, University of Pittsburgh, Pittsburgh, PA}


\begin{abstract}
\textbf{Fluids under nanoscale confinement differ -- and often dramatically -- from their bulk counterparts. A notorious feature of nanoconfined fluids is their inhomogeneous density profile along the confining dimension, which plays a key role in many fluid structural and transport phenomena in nanopores. Nearly five decades of theoretical efforts on predicting this phenomenon (fluid layering) have shown that its complexity resists purely analytical treatments; as a consequence, nearly all current approaches make extensive use of molecular simulations, and tend not to have generalizable predictive capabilities. In this work, we demonstrate that machine-learning-based models (in particular, a random forest model), trained upon large molecular simulation data sets, can serve as reliable surrogates in lieu of further molecular simulation. We show that this random forest model has excellent interpolative capabilities over a wide range of temperatures and confining lengthscales, and even has modest extrapolative ability. These results provide a promising pathway forward for developing models of nanoconfined fluid properties that are generalizable, lower cost than ``pure" molecular simulation, and sufficiently predictive for fluids-in-nanopores practitioners.}
\end{abstract}

\maketitle

\section{Introduction}

As compared to their bulk counterparts, fluids under nanoscale confinement are imbued with a host of remarkable properties \cite{NanofluidicsComingOfAge, ChengzhenNCF2020} -- including anomalous diffusion \cite{DiffusivityVsDensity,LayeringDependentDiffusion, ABF-NRA-DiffusionCNTs, FirstLayerDiffusivity,li2022produce}, viscosity \cite{NeekAmalViscosity}, thermal phenomena \cite{FLGKapitza, PoulikakosWaterThermalConductivity,NagoeCp,kim2008thermal}, pressure distribution \cite{ShiPressureCNT, ZouMarooPressure}, phase behavior \cite{Koga2000, TanakaBilayerIcePhase, TanakaWaterCNTPhaseDiagram, ABF-NRA-MultipleWaterPhases, AgrawalStranoWaterCNT}, contact-line dynamics \cite{LichterMovingContactLine}, and slip \cite{SendnerBocquet, BocquetInterlayerExchange, BarratBocquet_LargeSlipEffect, BocquetMicroNanoFlowReview, LichterFK, LichterFK_JFM, LichterSlipRateProcess}, amongst many others -- which all fundamentally stem from the comparability of an internal (molecular) lengthscale to the confinement lengthscale. At the heart of many of these anomalous properties is the inhomogeneous fluid density profile that emerges under confinement. Unlike bulk (single-phase) fluids, which are constant in density throughout space, nanoconfined fluids often feature significant density anisotropy, namely, long-ranged correlations along the direction of confinement. These correlations manifest as ordered \emph{layers} of fluid that run parallel to the confining boundary. There is a rich history of studying fluid layering phenomena at interfaces -- spanning nearly five decades -- using simulation \cite{ParsonageLayering, AbrahamLayering, ToxvaerdLayeringI, ToxvaerdLayeringII}, theory \cite{ToxvaerdLayeringII, OZInhomogeneous, RosenfeldFMT, StillingerBuffInhomogeneousFluids, ClusterExpansionInhomogeneousFluid, sisan2011end, AluruEQT, FluidsInCNTs, FirstLayerDensity}, and experiments \cite{YangZewail, Uhlig2019}. These studies have conclusively established the existence of fluid layering in a broad range of nanopore systems, characterized the lengthscales and density magnitudes associated with layering, and established connections between layering and other fluid properties under nanoconfinement. 
    
Despite this progress, one simple challenge continues to loom large: Given a specific choice of fluid and solid, and a set of thermodynamic conditions, can one produce an accurate estimate for the layered density profile, without resorting to time-intensive molecular simulations (or even more resource-intensive experimental characterization techniques)? On this question, we make two key observations:
\begin{enumerate}
    \item It is enormously difficult to develop a physically motivated, broadly generalizable, and predictive model for fluid layering. There are numerous approaches, grounded in rigorous theory, that (in theory) could be used to compute these density profiles; however, in practice, none of these approaches can generate a meaningful prediction using only simple details about the system setup. In particular, every existing theory contains at least one ``rug parameter" (so termed because a large amount of complex physics is ``swept under" this parameter). Such parameters are challenging (or impossible) to compute \emph{ab initio}; in fact, it is often the case that rug parameters cannot even be coarsely estimated without extensive amounts of simulation (followed by statistical fitting). For example, integral-equation methods (most notably, Ornstein-Zernike \cite{Hansen}) can in principle be used to calculate fluid density profiles in the presence of the external potentials induced by the confining boundaries \cite{OZInhomogeneous}; however, these methods require detailed \emph{a priori} knowledge of the homogeneous radial distribution function (not to mention a suitable closure relation). Approaches based upon (liquid-state) Density-Functional Theory, such as Fundamental Measure Theory \cite{RosenfeldFMT}, require ever-increasing numbers of non-obvious terms in the density functional (along with non-obvious weightings) in order to reproduce density profiles in systems of modest complexity. The Achilles heel of cluster-expansion techniques \cite{StillingerBuffInhomogeneousFluids, ClusterExpansionInhomogeneousFluid} is their reliance on virial coefficients. More recent ``semi-classical" theories \cite{AluruEQT, FirstLayerDensity} require detailed information about system-averaged energetics for fluid-solid and fluid-fluid interactions; this is occasionally feasible in simple geometries (employing hard-fought integrals), but not in general an analytically tractable demand. This limitation was clear even in the first-ever thermodynamic treatment of the layering phenomenon \cite{ToxvaerdLayeringII}, in which the Gibbs surface excess energy played the role of the rug. 
    \vspace{-2mm}
    \item There are significant practical benefits to simply knowing the inhomogeneous density profile. For example, knowledge of the layered fluid structure is valuable for the spatial calibration of two-viscosity models \cite{MyersTwoViscositySlip, ThomasMcGaugheyFastWaterTransport, MH-NRA-EnhancementFactor}, which describe a host of nanoconfined fluid transport phenomena as originating from two ``distinct" fluid phases adjoining each other within the system. Models for anomalous thermal properties of nanoconfined fluids (e.g. Kapitza resistance \cite{FLGKapitza} or heat capacity \cite{NagoeCp}) are often based upon knowledge of the vibrational modes supported by the interfacial fluid layer, which in turn requires knowledge of the density and spatial extent of this layer. Approximate knowledge of the equilibrium nanoconfined fluid density profile can also accelerate molecular simulations, by providing initial conditions that require less evolution in order to attain a state of statistical equilibrium. In other words, even in the absence of any first-principles rationalization, there are numerous applications that would benefit from rapid and accurate estimates for nanoconfined fluid density profiles.
\end{enumerate}

Based upon these observations, it is clear that there is a significant need for fast, predictive tools that describe nanoconfined fluid structure; this need is unmet (and unlikely to be met) by existing methods. The premise for the present work naturally follows: A machine-learned (ML) surrogate model for fluid layering, trained using an extensive data set from molecular simulations of confined fluids under a wide variety of degrees of confinement and thermodynamic conditions, has tremendous practical value. We have developed several such ML models and, as we describe below, we find that these models, once trained, are capable of producing accurate predictions for fluid-layering phenomena (and at negligible computational cost as compared to the underlying molecular simulations). Moreover, although our models (like many ML models) are not fundamentally grounded in physical principles, we demonstrate that these models respect relevant physical constraints. Unlike the existing approaches detailed above, we are optimistic that ML-driven models have a clear pathway forward for solving the core challenge of predicting fluid density profiles in a broad range of nanoconfined systems.

\section{Methods}
In the discussion that follows, we focus our efforts on modeling the density profiles of water confined within graphene nano-slits. Predicting these density profiles is a notoriously challenging problem, especially in the narrowest slits that accommodate water imbibition (0.4 nm); this is both because the number of fluid layers in the channel can change suddenly as the channel increases in width, and also because the layered structures induced by each confining boundary can significantly overlap each other. Throughout this work, we refer to the system temperature as $T$ and the nano-slit width as $w$.

\subsection{Molecular-Dynamics Simulations}
Using molecular-dynamics (MD) simulations, performed using LAMMPS\cite{lammps}, we have computed density profiles for 2160 systems of nanoconfined water. In these simulations, water molecules are modeled using the SPC/E potential \cite{SPCE}. Oxygen-carbon interactions are modeled using a Lennard-Jones potential, with $\sigma_\text{O-C}= 3.28$ \AA{} and $\varepsilon_\text{O-C} = 0.114$ kCal/mol. Water molecules are confined within a planar domain by rigid graphene walls (Fig.~\ref{fig:workflow}a). The cross-sectional area of the system is 30 \AA{} by 30 \AA{}, with periodic boundaries applied in each dimension. We systematically vary the nano-slit width $w$ between 4 \AA{} and 75 \AA{}. In each system, we insert a number of water molecules consistent with a system-averaged density of 1 g/cm$^3$ (for the purpose of computing system-averaged density, recognizing that there is a fluid-solid stand-off distance at each interface, we treat the \emph{effective} width of the channel as $w - \sigma_\text{O-C}$).

We employ a timestep of 2 fs. Using the Nosé-Hoover thermostat \cite{nose,hoover}, each simulation was run in the canonical ensemble for 200 ps to attain a target temperature, $270\text{ K}\leq T\leq 328\text{ K}$. The system is then switched to the microcanonical ensemble, run for an additional 200 ps to reach statistical equilibrium, after which we begin a 0.5 ns sampling period. When sampling densities, we used fixed bin widths (0.1 \AA{}) along the confining dimension ($z$-axis), and so the output vectors range in size from 40 elements to 750 elements. To create a data set with uniform size along the spatial dimension, amenable for the ML models described below, we zero-pad the water density profiles symmetrically on both sides, to a constant length of 1000 elements. 

\subsection{Machine Learning}
\begin{figure}[]
  \centering
  \includegraphics[width=\linewidth]{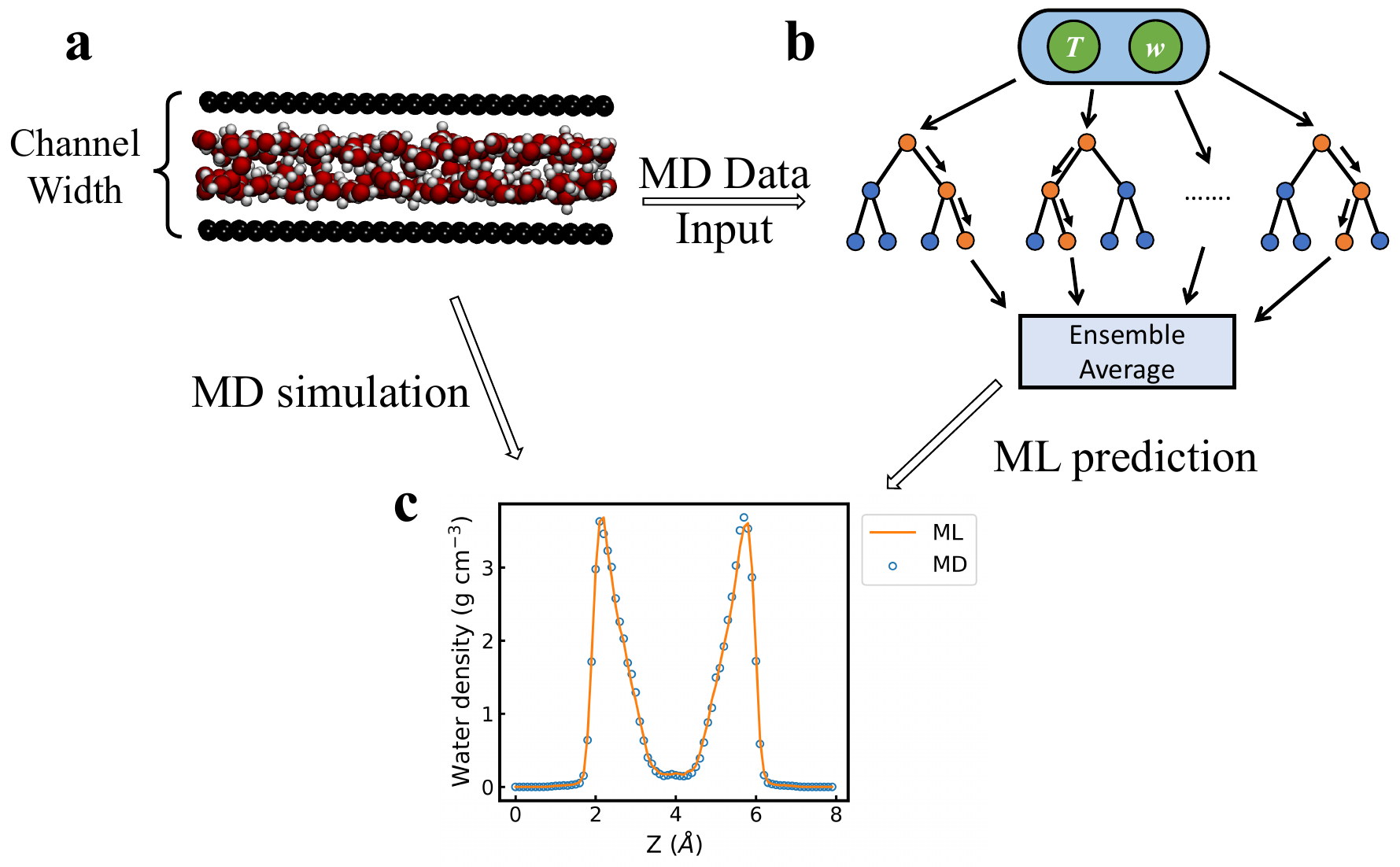}
\caption{(a) Schematic of MD setup for water molecules confined by graphene walls ($w= 10 \textrm{\AA} $). (b) Schematic of a random forest algorithm that maps inputs of temperatures and channel widths to outputs of water density profiles. (c) Comparison between water density profiles from MD and predictions from a trained random forest.}
\label{fig:workflow}
\end{figure}

Having generated the MD data set, we have trained machine learning (ML) models to predict the water density profile, using temperature and channel width as inputs (Fig.~\ref{fig:workflow}b). 
We train and compare three ML models: Random forest (RF), XGBoost, and multilayer perceptron (MLP). For all three ML models, the choice of loss function is the mean-squared error (MSE) between the MD and ML results. 
The random forest and XGBoost algorithms are carried out using Scikit-Learn\cite{scikit-learn} and XGBoost-python\cite{chen2016xgboost} (default parameters), respectively. Our MLP is a fully connected neural network, implemented with the Pytorch\cite{paszke2019pytorch} package. The network consists of 6 layers (5 hidden and 1 output), with 64, 128, 256, 512, and 1024 neurons for the hidden layers and 1000 neurons for the output layer. To regularize the neural network and prevent overfitting, dropout\cite{srivastava2014dropout} is applied after the third, fourth, and fifth hidden layers with dropout rates of 0.05, 0.2, 0.3 for each layer, respectively. A ReLU non-linear activation function is applied after each hidden layer. Softplus activation is applied after the output layer, which not only smooths the predicted profiles but (critically, from a physical perspective) ensures non-negativity of the output densities. For the density profiles obtained via the MLP, we perform an additional post-processing step to renormalize the profile such that total fluid mass is conserved as compared to the MD simulations at that channel width; intriguingly, for random forest and XGBoost, this step was not necessary for any of the systems considered (deviations from mass conservation were found to be on the order of $10^{-3}$ or smaller in relative magnitude).

The MLP is trained over 600 epochs with the Adam optimizer\cite{kingma2014adam} and an initial learning rate of $10^{-3}$, which decays to $10^{-4}$ after 360 epochs. We separate the data set into two components: One for assessing the strength of our models for interpolation, and the other for assessing its potential for extrapolation (in both cases, with respect to temperature). The temperature interpolation part spans $\SI{280}{\kelvin}\leq T \leq \SI{300}{\kelvin}$ , while the extrapolation part spans $\SI{270}{\kelvin} \leq T \leq \SI{278}{\kelvin} $ and $\SI{322}{\kelvin} \leq T \leq \SI{328}{\kelvin} $. The training data is selected from a halo-shaped region (Fig.~\ref{fig:MSE_heatmap}a), within which we select 80\% of the systems for training. The remaining 20\% of systems within this halo-shaped region, and the interior of the halo, are reserved for assessing interpolation quality. We have checked our results with 5-fold cross validation (the interior of the halo is always reserved for testing interpolation). In sum total, there are 848, 664, and 648 density profiles in the training, test-interpolation, and test-extrapolation sets, respectively.

\section{Results and Discussion}

\begin{figure}[hbt!]
  \centering
  \includegraphics[width=\linewidth]{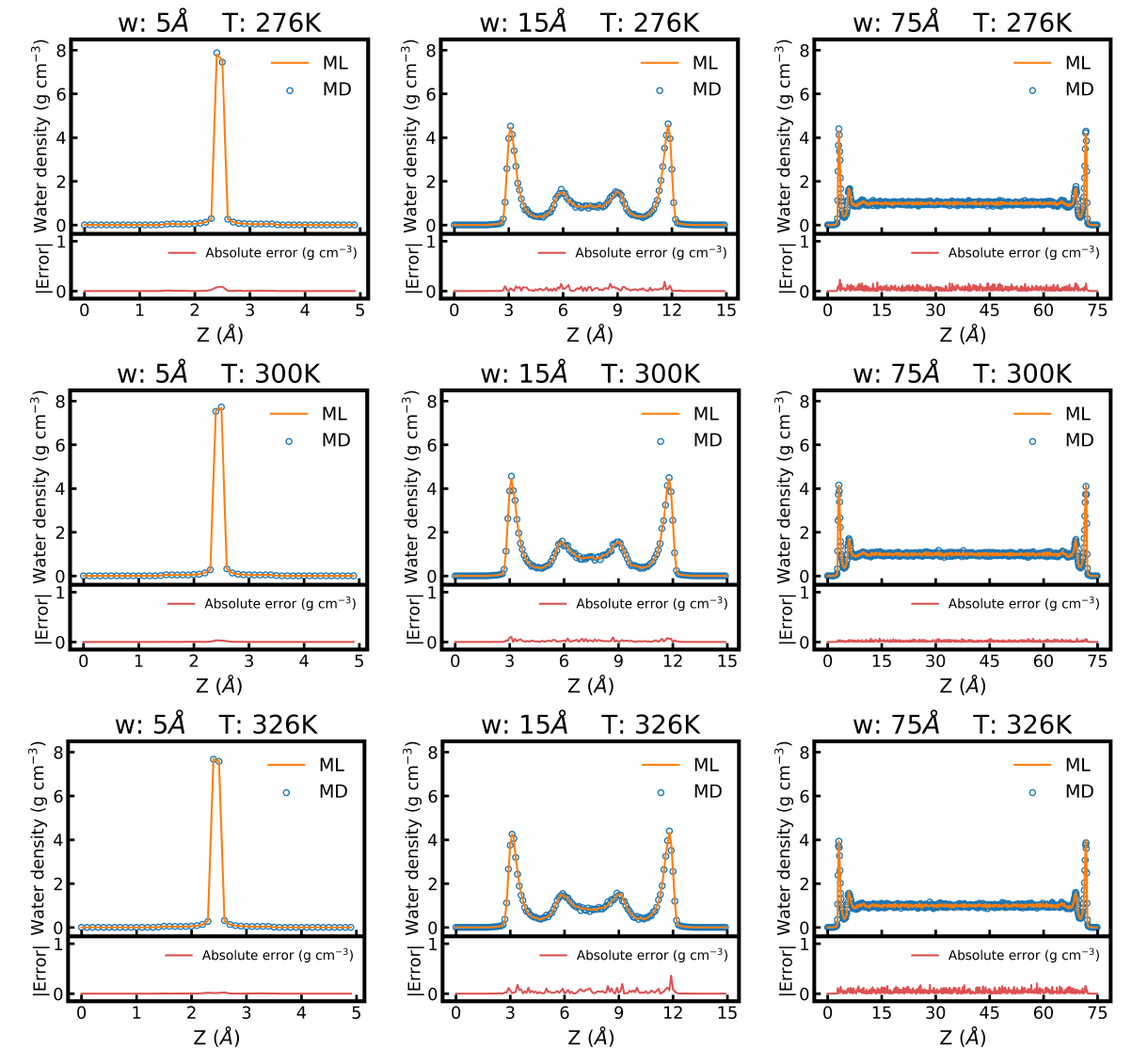}
\caption{Comparison between the ML (random forest) and MD water density profiles for $T\in \{276, 300, 326\}$ K and $w\in \{5, 15, 75\}$ \textrm{\AA}. The red curve below each density profile shows the absolute error between the ML and MD results.}
\label{fig:density_profile}
\end{figure}

We concentrate our discussion on the results of the random forest model; we discuss the performance of XGBoost and MLP at the end of this section. We find that the water density profiles predicted by the random forest model agree closely with the MD results across the range of temperatures and channel widths studied, with absolute errors that are at most on the order of 0.1 g/cm$^3$, and generally much less (Fig.~\ref{fig:density_profile}). This strong agreement evidences our core claim, which is that -- as in numerous other fields -- ML models can play a critical role in nanoconfined fluids research by serving as surrogates for a computationally intensive task. Within the errors, we observe a systematic trend: The largest magnitudes of error occur at or in the vicinity of the locations of peak density (the layers of fluid adsorbed at the solid boundaries). In other words, and in alignment with reasonable physical expectations, these data-driven models struggle the most in precisely matching the features of the density profile for which there is the least data; in contradistinction, the training set features numerous examples of density profiles with well established and spatially extensive bulk regions, which are well captured by the random forest model. It is worth noting that this is a systematic challenge for any purely data-driven approach: For any fixed set of simulations, broadly representative of a nanopore-confined fluid system, there will almost certainly be \emph{less} data available to fully characterize the peaks as compared to the bulk (which of course does not require ML to characterize). ML is not a silver bullet for this problem.

Nevertheless, and reassuringly, we emphasize that these discrepancies are small in magnitude. Specifically, they are small enough for ML models to be useful as surrogates in the applications discussed above. To provide concrete context, in our nano-slit geometry, a discrepancy within a single sampling bin that is on the order of 0.1 g/cm$^3$ translates to less than half a water molecule within that bin, which is clearly well beyond the precision that could be required by any theoretical or experimental application of these density profiles. In fact, for the levels of agreement we observe between MD and the random forest model trained on MD, it is almost certain that uncertainties in the underlying MD simulations (associated, e.g., with statistical sampling) easily dwarf any discrepancies introduced by the ML model. In other words, for this problem, ML faithfully serves its purpose as a reliable surrogate. Unsurprisingly, although worthy of emphasis, is that the ML model, once trained, can be evaluated at significantly lower cost than the full MD simulation. To contextualize this claim, we note that whereas the MD simulation for any given system would require at least $\mathcal O(10^3)$ seconds, a single evaluation of the random forest model can be carried out in $\mathcal O(10^{-3})$ seconds on the same machine (and, for full context, with the training data set in hand, the training procedure itself requires $\mathcal O(10^{1})$ seconds). 

\begin{figure}[]
    \centering
    \includegraphics[width=\textwidth, height=0.92\textheight]{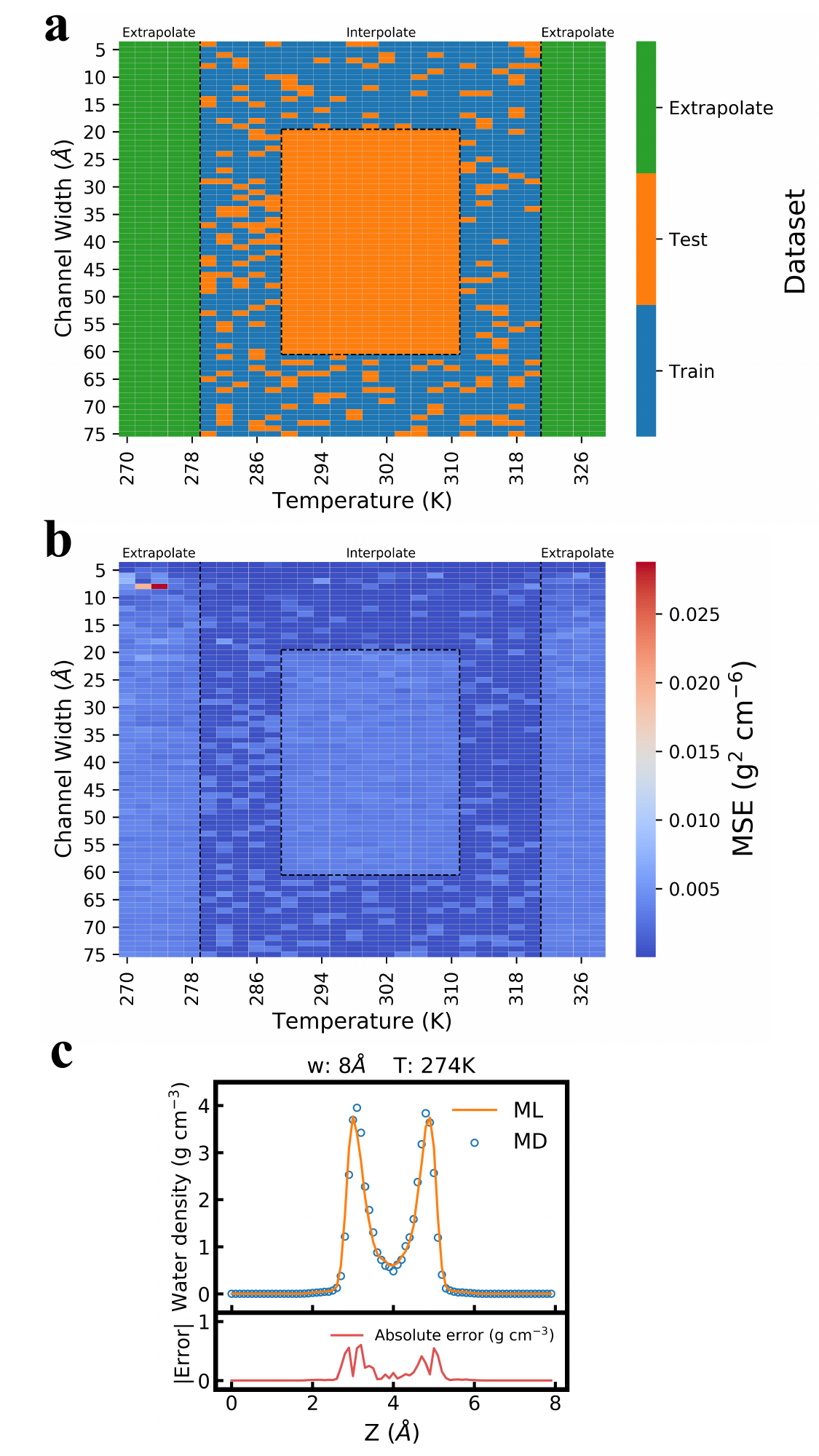} 
\caption{(a) Schematic illustrating the partitioning of our data set into three components: Blue cells denote the training set, orange cells denote the test-interpolation set, and green cells denote the test-extrapolation set. (b) Heat map of the MSE between ML and MD water density profiles. Darker blue represents lower MSE, while darker red represents higher MSE. (c) Water density profile for the system with the single highest MSE between ML and MD ($T=274$ K, $w=8 $ \AA{}), which nevertheless features reasonable agreement.}
\label{fig:MSE_heatmap}
\end{figure}

Especially tantalizing, we find that the random forest model is even capable of modest extrapolation. Within a window of approximately 10 K of the edges of the interpolation region (Fig.~\ref{fig:MSE_heatmap}b), and across almost all channel widths studied, the random forest model is able to yield density profiles whose typical errors are comparable to the errors in the interpolation region. Although this result, on its face, runs counter to the conventional wisdom that ML performs poorly on extrapolation tasks, it is nevertheless reasonable: Our fluid density profiles, despite nominally living in a high-dimensional space, are ultimately much lower-dimensional creatures. They follow reliable (if intricate) patterns, and are almost fully characterized by a small handful of quantities, including the fluid-solid standoff distance \cite{FluidsInCNTs}, the width of the first fluid layer, and the total fluid content in the first fluid layer \cite{FirstLayerDensity}. Our extrapolation results bear promise that ML-driven models can be especially useful for modeling fluid layering phenomena in nanopores with generality. Reassuringly, the most significant extrapolation errors made by the ML model occur in the northwest corner of Fig.~\ref{fig:MSE_heatmap}b, where the lowest temperatures and smallest confining lengthscales that we studied are most likely to drive a phase transition \cite{PhaseDiagramConfinedWater}. This hypothesis is supported by the finding that, in this corner, the ML \emph{underpredicts} the size of the peaks seen in MD (Fig.~\ref{fig:MSE_heatmap}c); in view of the Hansen-Verlet freezing criterion \cite{HansenVerletFreezing}, the MD exhibits stronger characteristics of freezing than the ML model predicts. This observation underscores the critical need, when deploying such data-driven methods, to ensure that the training set contains sufficient data in every phase of interest.

\begin{figure}[]
    \centering
    \includegraphics[width=\linewidth]{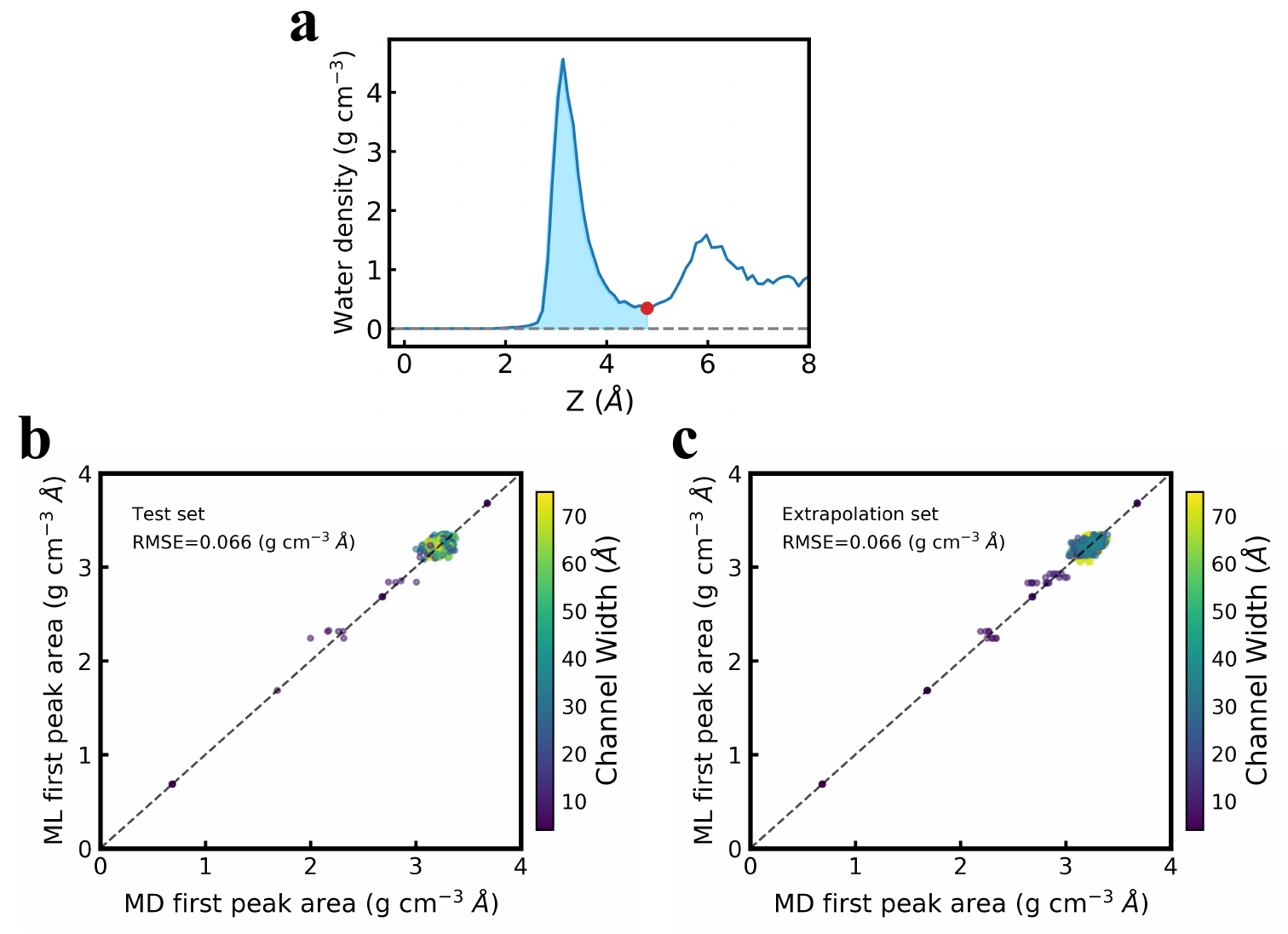}
\caption{(a) Area below the first peak (light blue colored region) represents water adsorbed at a graphene boundary. (b) and (c) Parity plots for the test-interpolation set and test-extrapolation sets, respectively. The black dashed line indicates parity. Data points are colored according to channel width.}
\label{fig:peakArea}
\end{figure}

To further evidence our claim that ML models can serve as meaningful surrogates, we explore their capability to predict the area under the first peak of the water density profile (Fig.~\ref{fig:peakArea}a). This quantity, which describes fluid adsorbed at the solid interface, is a key measure of fluid wetting, and plays a critical role in a wide variety of fluids-in-nanopores applications. We assess the first peak area obtained via ML as compared to MD in both the test-interpolation and test-extrapolation sets, and represent these results on parity plots in (Fig.~\ref{fig:peakArea}b-c). We observe that the overall root-mean-squared-error (RMSE) of the first peak area is 0.07 g$\cdot \textrm{\AA}/$cm$^{3}$ for both test-interpolation and test-extrapolation sets, with all systems falling in close proximity to the line of parity.

\begin{figure}[hbt!]
    \centering
    \includegraphics[width=\linewidth]{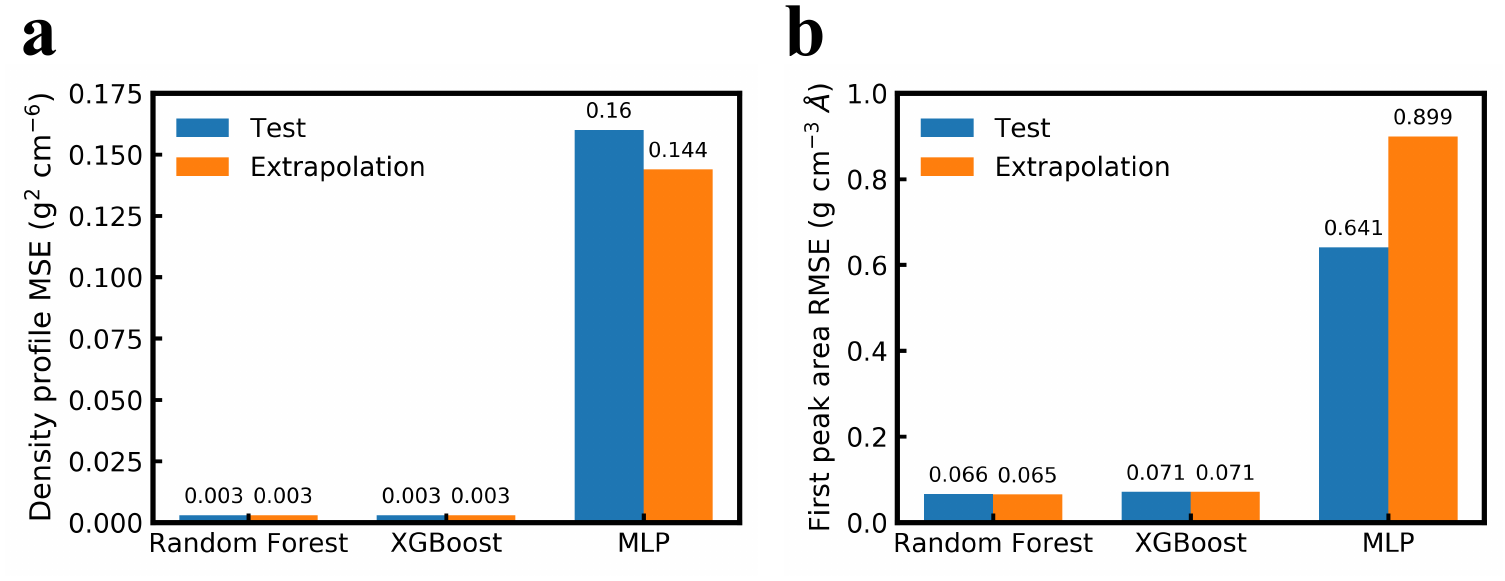}
\caption{Comparison between the accuracy of three ML models based upon two metrics: (a) data-set-averaged water density profile MSE and (b) data-set-averaged RMSE of first peak area of samples in the test-interpolation and test-extrapolation sets.}
\label{fig:modelComp}
\end{figure}

Finally, we comment on the performance of random forest vis-\`a-vis XGBoost and MLP. The data-set-averaged MSE of water density profiles and RMSE of the first peak area are shown in (Fig.~\ref{fig:modelComp}a) and (Fig.~\ref{fig:modelComp}b), respectively. Random forest and XGBoost have comparable data-set-averaged MSEs for their water density profile predictions ($\sim 3\cdot 10^{-3}($g/cm$^{3}$)$^2$ for both the test-interpolation and test-extrapolation sets). These two methods have MSEs that are $\sim$98$\%$ lower than MLP's data-set-averaged MSE, for both the test-interpolation and test-extrapolation sets. Random forest exhibits the lowest RMSE for area of the first peak ($\sim$0.07 g$\cdot \textrm{\AA} /$cm$^{3}$ for both the test-interpolation and test-extrapolation sets), which is 7$\%$ and 8$\%$ lower error than for XGBoost, and 90$\%$ and 93$\%$ lower error than MLP for the test-interpolation and test-extrapolation sets, respectively. As a final comment, we note that by producing the most accurate predictions for all of the fluid density measures we studied -- in addition to ease of use, flexibility, and the capability for feature ranking -- distinguish the random forest model as the most promising ML approach of the ones that we studied.

\section{Conclusion}
We have shown that ML models, trained upon MD simulations, are capable of acting as reliable surrogates in lieu of further (costly) MD simulations. In particular, we demonstrate that several ML models (engineered to respect basic physical constraints, including non-negativity of densities and conservation of mass) are capable of mapping inputs of system temperature and confining lengthscale to accurate predictions of the density profile of water nanoconfined within a graphene nano-slit, both in an interpolative and a (modestly) extrapolative capacity. In particular, our random forest model can be evaluated approximately six orders of magnitude faster than running a new MD simulation, and thus has considerable utility as a surrogate model. Although these models are certainly not perfect (e.g. in the vicinity of a phase boundary), we conclude that they are a promising tool for predicting fluid density profiles under nanoconfinement. Indeed, from the perspective of a fluids-in-nanopores ``practitioner," who above all else needs fast and reliable estimates of nanoconfined fluid densities, a little machine learning goes a long way.
\newline

\bibliographystyle{naturemag_noURL}
\bibliography{main}

\subsection *{Competing interests.}
The authors declare no competing interests.

\subsection *{Correspondence.}
Correspondence and requests for materials should be addressed to yuanhaol@alumni.cmu.edu.


\subsection *{Data Availability.}
The data that support the findings of this study are available from the corresponding author upon reasonable request.

\subsection *{Code Availability.}
All the codes used in this study are available from the corresponding author upon reasonable request.

\balance

\renewcommand{\figurename}{{\bf Extended Data Fig.}}
\setcounter{figure}{0}

\balance
\end{document}